\documentstyle[amssymb,12pt]{article}
\begin{document}

\title{On violation of the Pauli principle and dimension of the Universe at very
large distances}
\author{A. A. Kirillov \\
%EndAName
{\em Institute for Applied Mathematics and Cybernetics} \\
{\em 10 Uljanova Str., Nizhny Novgorod, 603005, Russia} \\
e-mail: kirillov@unn.ac.ru}
\date{}
\maketitle

\begin{abstract}
We discuss a Modified Field theory (MOFT) in which the number of fields can
vary. It is shown that in MOFT fermions obey the so-called parastatistics in
which the order of the parastatistics depends on scales. In particular, at
very large distances $r>r_{\min }$ fermions violate the Pauli principle. It
is also shown that in MOFT in some range of scales $r_{\max }>r>r_{\min }$
the Universe acquires features of a two-dimensional space whose distribution
in the observed 3-dimensional volume has an irregular character. This
provides a natural explanation to the observed fractal distribution of
galaxies and the logarithmic behavior of the Newton's potential for a point
source.
\end{abstract}

\pagebreak

\section{Introduction}

The Modified Field Theory (MOFT) was suggested in Ref. \cite{K99} to account
for spacetime foam effects, and it was recently demonstrated in Ref. \cite
{KT02} that MOFT provides a reasonable alternative to the dark matter
paradigm. It was shown that MOFT possesses a nontrivial \ vacuum state which
leads to a scale-dependent renormalization of all interaction constants $%
\alpha ^{2}\rightarrow $ $\alpha ^{2}N\left( k\right) $ (where $k$ is the
wave number and $\alpha $ is either the electron charge $e$, the gauge
charge $g$ , or the gravitational constant $\sqrt{G}$) with some structural
function $N\left( k\right) $ which reflects the topology of the momentum
space (for details we send readers to Ref. \cite{KT02}). This means that
particles lose their point-like character in MOFT and acquire a specific
distribution in space, i.e. each point source is surrounded with a dark halo
which carries charges of all sorts. In the simplest case properties of the
vacuum and that of the dark halos are described by two characteristic scales 
$r_{\min }$ and $r_{\max }$ between which interaction constants increase $%
\alpha ^{2}\left( \ell \right) \sim \alpha _{0}^{2}\ell /r_{\min }$ ($\ell
=2\pi /k$) and the Newton's and Coulomb's interaction energies show the
logarithmic $V\sim \ln r$ (instead of $1/r$) behavior. \ Thus, MOFT
reproduces the basic phenomenological feature of the Modified Newtonian
Dynamics (MOND) proposed by Milgrom \cite{mil} to model the dark matter
effects (see also the list of references devoted to MOND at the site \cite
{Site}) but conceptually MOFT differs from MOND very much.

In addition to logarithmic potentials, MOFT seems to require that the
Universe has a fractal structure in the same range of scales with dimension $%
D\approx 2$ \cite{K02}, which is supported by the observed fractal
distribution of galaxies (e.g., see Refs. \cite{Ruf} - \cite{R99}). In
particular, the fractal distribution of luminous matter is the only possible
picture of the Universe which appears in the case when the maximal scale is
absent (i.e., $r_{\max }\rightarrow \infty $). At the same time, it is shown
in Ref. \cite{K02} that the theoretical scheme of MOFT provides a full
agreement of the observed fractal distribution of luminous matter with
homogeneity of the Universe and observational limits on $\Delta T/T$ in the
microwave background.

However, in the present form MOFT is not a self-consistent theory,
regardless of the good features pointed out. The problem is that\ in MOFT
fermions and bosons are described in essentially different ways, for MOFT
was originally developed only for bosonic fields \cite{K99}. In fact, there
exists an argument which shows that the situation has to be somewhat
different. Indeed, the basic idea of MOFT is that the nontrivial topology of
space displays itself in the multivalued nature of all observable fields,
i.e. the number of fields represents an additional dynamical variable which
was shown to depend on the position in the momentum space \cite{K99,KT02}.
The topological origin of this additional variable means that the number of
fields $N\left( k\right) $ should represent a certain function of the wave
number $k$, which must be the same for all types of fields and which serves
as a geometric characteristic of the momentum space. In fact, this function
defines the spectral number of modes in the interval between $k$ and $k+dk$ 
\begin{equation}
g\frac{k^{3}N\left( k\right) }{4\pi ^{2}}\frac{dk}{k},  \label{mes}
\end{equation}
where $g$ is the number of polarization states. Since $N\left( k\right) $
has a pure geometrical nature, it is clear that the measure (\ref{mes})
characterizes the number of degrees of freedom of a particular point
particle and must hold for all types of particles, regardless of their
statistics, contrary to the claim of Ref. \cite{K99} that the modification
should involve only bosonic fields. In fact, a self-consistent consideration
of fermions also requires a description in terms of variable number of
fields.

In the present paper we suggest an extension of MOFT to the case of
fermionic fields as well. From the mathematical standpoint the
generalization of the quantum field theory to the case of a variable number
of fields does not depend on the statistics of particles and remains the
same as in the case of bosonic fields (e.g., see Ref. \cite{K99} ).
Fermions, however, involve two modifications. First, the number of particles
in every mode takes now only two values $n=0,1$ (according to the Fermi
statistics) and, secondly, the fundamental operators of creation and
annihilation of fermionic modes have to obey the Bose statistics (otherwise
the total number of modes will be always restricted by $N\left( k\right)
\leq 2$ for all $k$). At very large scales corresponding to $k<2\pi /r_{\min
}$, the number of modes $N\left( k\right) $ was shown to take values $%
N\left( k\right) >1$ \cite{KT02} and, therefore, the Pauli principle is
violated (up to $N\left( k\right) $ fermions can occupy the same quantum
state which corresponds to a wave number $k$). Thus, we see that particles
obey in MOFT the so-called parastatistics (or the Green statistics) where
the function $N\left( k\right) $ plays the role of the order of the
parastatistics. We note that the parastatistics was first suggested by Green
in Ref. \cite{G53} and it has been studied in many papers since then (e.g.,
see Refs. \cite{Vol} - \cite{gov} and, for more recent discussions, Refs. 
\cite{conf}). In this manner, we see that MOFT gives a specific realization
of the parafield theory in which the order of the statistics represents a
new variable related to the topology of space.

The fact that the number of degrees of freedom for any point particle is
described by the measure (\ref{mes}) means that the number of fields $%
N\left( k\right) $ characterizes the density of physical space. This allows
us to give a common, pure geometrical interpretation to both observational
phenomena: logarithmic potentials and the fractal distribution of galaxies,
as we briefly demonstrate in the present paper. Namely, we show that in the
range of scales $r_{\min }<r<r_{\max }$ some kind of reduction of the
dimension of space happens, i.e. an effective dimension of the Universe is $%
D\approx 2$.

\section{Description of particles in MOFT}

Let ${\it \psi }$ be an arbitrary field which, upon the expansion in modes,
is described by a set of creation and annihilation operators $\left\{
a_{\alpha ,k},a_{\alpha ,k}^{+}\right\} $, where the index $\alpha $
numerates polarizations and distinguishes between particles and
antiparticles. In what follows, for the sake of simplicity, we ignore the
presence of the additional discrete index $\alpha $. These operators are
supposed to satisfy the relations 
\begin{equation}
a_{k}a_{p}^{+}\pm a_{p}^{+}a_{k}=\delta _{kp},  \label{b}
\end{equation}
where the sign $\pm $ depends on the statistics of particles. The expression
(\ref{mes}) implies that in MOFT the number of fields is a variable (which
must be the same for all types of particles) and, therefore, the set of
field operators $\left\{ a_{k},a_{k}^{+}\right\} $ is replaced with \ the
expanded set $\left\{ a_{k}\left( j\right) ,a_{k}^{+}\left( j\right)
\right\} ,$ where $j\in \left[ 1,...,N\left( k\right) \right] $. For a free
field, the energy is an additive quantity, so it can be written as 
\begin{equation}
H_{0}=\sum_{k}\sum_{j=1}^{N\left( k\right) }\omega _{k}a_{k}^{+}\left(
j\right) a_{k}\left( j\right) ,  \label{en}
\end{equation}
where $\omega _{k}=\sqrt{k^{2}+m^{2}}$. In the general case, the total
Hamiltonian $H=H_{0}+V$ (where the potential term $V$ is responsible for
interactions) can be expanded in the set of operators (e.g., see for details
Ref. \cite{K99}) 
\begin{equation}
A_{m,n}\left( k\right) =\sum_{j=1}^{N\left( k\right) }\left( a_{k}^{+}\left(
j\right) \right) ^{m}\left( a_{k}\left( j\right) \right) ^{n}.  \label{a}
\end{equation}

In a complete theory with a variable number of fields, quantum states can be
classified by occupation numbers. To this end, we consider the set of
creation and annihilation operators for field modes $\left\{ C\left(
n,k\right) ,C^{+}\left( n,k\right) \right\} $, where $k$ is the wave number
and $n$ is the number of particles in the given mode. These operators obey
the standard relations 
\begin{equation}
C\left( n,k\right) C^{+}\left( m,k^{\prime }\right) \mp C^{+}\left(
m,k^{\prime }\right) C\left( n,k\right) =\delta _{nm}\delta _{kk^{\prime
}}\;.  \label{c}
\end{equation}
and should be used to construct the Fock space in MOFT. The sign $\mp $ in ( 
\ref{c}) depends on the symmetry of the wave function under field
permutations, i.e., on the statistics of fields. In the case of the
so-called parafield theory there exists a spin-statistics theorem \cite{gov}
(analogous to the Pauli theorem) which states that bosonic modes should be
quantized according to the Fermi statistics, while fermionic modes should
obey the Bose statistics. Since MOFT represents a specific generalization of
the parafield theory, we will assume the same rule for MOFT.

In terms of $C$ and $C^{+}$, operators (\ref{a}) can be expressed as follows 
\begin{equation}
\widehat{A}_{m_{1},m_{2}}\left( k\right) =\sum\limits_{n}\frac{\sqrt{\left(
n+m_{1}\right) !\left( n+m_{2}\right) !}}{n!}C^{+}\left( n+m_{1},k\right)
C\left( n+m_{2},k\right)  \label{A}
\end{equation}
where the sum is taken over the values $n=0,1,...$ $\ $ in the case of
bosons, and $n=0,1$ in the case of fermions. Thus, the eigenvalues of the
Hamiltonian of a free field take the form 
\begin{equation}
\widehat{H}_{0}=\sum_{k}\omega _{k}\widehat{A}_{1,1}\left( k\right)
=\sum_{k,n}n\omega _{k}N\left( n,k\right) ,
\end{equation}
where $N\left( n,k\right) $ is the number of modes for fixed values of the
wave number $k$ and the number of particles $n$ (i.e., $N\left( n,k\right)
=C^{+}\left( n,k\right) C\left( n,k\right) $).

Thus, the field state vector $\Phi $ is a function of the occupation numbers 
$\Phi \left( N\left( n,k\right) ,t\right) $, and its evolution is described
by the Shr\"{o}dinger equation 
\begin{equation}
i\partial _{t}\Phi =H\Phi .
\end{equation}
Consider the operator 
\begin{equation}
N\left( k\right) =\widehat{A}_{0,0}\left( k\right)
=\sum\limits_{n}C^{+}\left( n,k\right) C\left( n,k\right)  \label{N(k)}
\end{equation}
which characterizes the total number of modes for a fixed wave number $k$.
In standard processes when the number of fields is conserved (e.g., when
topology transformations are suppressed) $N\left( k\right) $ is a constant
of the motion $\left[ N\left( k\right) ,H\right] =0$ and, therefore, this
operator can be considered as an ordinary fixed function of wave numbers.

\ Consider now the particle creation and annihilation operators. Among the
operators $\widehat{A}_{n,m}\left( k\right) $ are some which change the
number of particles by one 
\begin{equation}
b_{m}^{-}\left( k\right) =\widehat{A}_{m,m+1}\left( k\right)
,\;\;b_{m}^{+}\left( k\right) =\widehat{A}_{m+1,m}\left( k\right) ,
\end{equation}
and which replace the standard operators of creation and annihilation of
particles, i.e., they satisfy the relations 
\begin{equation}
\left[ \widehat{n},b_{m}^{(\pm )}\left( k\right) \right] =\pm b_{m}^{(\pm
)}\left( k\right) ,\;\;\;\left[ H_{0},b_{m}^{(\pm )}\left( k\right) \right]
=\pm \omega _{k}b_{m}^{(\pm )}\left( k\right) ,
\end{equation}
where 
\begin{equation}
\widehat{n}=\sum_{k}\widehat{n}_{k}=\sum_{k,n}nN\left( n,k\right) .
\end{equation}
In the case of fermions there exist only two such operators $b_{0}^{+}\left(
k\right) $ and $b_{0}^{-}\left( k\right) $, while in the case of bosons the
total number of creation/annihilation operators is determined by the
structure of the interaction term $V$. In the simplest case (e.g., in the
electrodynamics) the interaction term is expressed solely\ via $%
b_{0}^{+}\left( k\right) $ and $b_{0}^{-}\left( k\right) $ and, therefore,
we can introduce creation/annihilation operators for the effective field 
\begin{equation}
\widetilde{a}_{k}=\frac{1}{\sqrt{N\left( k\right) }}b_{0}^{-}\left( k\right)
,\;\;\widetilde{a}_{k}^{+}=\frac{1}{\sqrt{N\left( k\right) }}b_{0}^{+}\left(
k\right) ,  \label{ef}
\end{equation}
which satisfy the standard commutation relations, i.e., $\left[ \widetilde{a}
_{k},\widetilde{a}_{p}^{+}\right] =\delta _{kp}$. This restores the standard
theory but new features, however, appear. First, the renormalization (\ref
{ef}) results in the renormalization of interaction constants (e.g., see for
details \cite{KT02}) and, secondly, in the region of wave numbers in which $%
N\left( k\right) >1$ fermions violate the Pauli principle (up to $N\left(
k\right) $ fermions can be created at the given wave number $k$).

\section{Vacuum state in MOFT}

In this section we describe the structure of the vacuum state for bosons and
fermions. The true vacuum state in MOFT is defined by \ the relation 
\begin{equation}
C\left( n,k\right) \left| 0\right\rangle =0.
\end{equation}
In the true vacuum state all modes are absent $N\left( k\right) =0$, hence
no particles can be created and all observables related to the field are
absent. Thus, the true vacuum state corresponds to the absence of physical
space and, in reality, cannot be achieved. Assuming that upon the quantum
period of the evolution of the Universe topology transformations are
suppressed, we should require that the number of fields conserves $N\left(
k\right) =const$ in every mode. Then we can define the ground state of the
field ${\it \psi }$ (which is the vacuum from the standpoint of particles)
which is the vector $\Phi _{0}$ satisfying the relations 
\begin{equation}
b_{m}\left( k\right) \Phi _{0}=0  \label{0}
\end{equation}
for all values $k$ and $m=0,1,...$ . However, these relations still do not
define a unique ground state and should be completed by relations which
specify the distribution of modes $N\left( k\right) $. In the case of
bosons, additional relations can be taken in the form 
\begin{equation}
b_{N\left( k\right) +m}^{+}\left( k\right) \Phi _{0}=0
\end{equation}
where $m=0,1,...$. Then the state $\Phi _{0}$ corresponds to the minimum
energy for a fixed mode distribution $N\left( k\right) $ and is
characterized by the occupation numbers 
\begin{equation}
N\left( n,k\right) =\theta \left( \mu _{k}-n\omega _{k}\right)  \label{d}
\end{equation}
where $\theta \left( x\right) $ is the Heaviside step function and $\mu _{k}$
is the chemical potential which can be expressed via the number of modes $%
N\left( k\right) $ as 
\begin{equation}
N\left( k\right) =\sum_{n}\theta \left( \mu _{k}-n\omega _{k}\right) =1+ 
\left[ \frac{\mu _{k}}{\omega _{k}}\right] .
\end{equation}
In particular, from (\ref{d}) we find that the ground state for bosonic
fields contains real particles 
\begin{equation}
\widetilde{n}_{k}=\sum_{n=0}^{\infty }nN\left( n,k\right) =\frac{1}{2}
N\left( k\right) \left( N\left( k\right) +1\right)  \label{hp}
\end{equation}
and, therefore, corresponds to a finite energy $E_{0}=\sum \omega _{k} 
\widetilde{n}_{k}$. These particles, however, are ''dark'' for they
correspond to the ground state.

In the case of fermions the only nontrivial operator $b_{m}^{\left(
\pm\right) }\left( k\right)$ corresponds to $m=0$, and the additional
relations read

\begin{equation}
\left( b_{0}^{+}\left( k\right) \right) ^{N\left( k\right) +1}\Phi _{0}=0.
\label{r}
\end{equation}
These relations mean that the total number of particles $n_{k}$ which can be
created at the given wave number $k$ takes values $n_{k}=0,1,...,N\left(
k\right) $ (i.e., it cannot exceed the number of modes). In this case the
ground state $\Phi _{0}$ corresponding to the fixed mode distribution $%
N\left( k\right) $ is characterized by the occupation numbers 
\begin{equation}
N\left( n,k\right) =N\left( k\right) \delta _{n,0}.  \label{f}
\end{equation}
In contrast to the case of bosonic fields, the ground state (\ref{f})
contains no particles and corresponds to the zero energy $E_{0}=0$. Formally
the ground state (\ref{f}) can be constructed from the true vacuum state as
follows 
\begin{equation}
\Phi _{0}=\left| N\left( k\right) ,0\right\rangle =\prod_{k}\frac{\left(
C^{+}\left( 0,k\right) \right) ^{N\left( k\right) }}{\sqrt{N\left( k\right)
! }}\left| 0\right\rangle ,
\end{equation}
while the basis of the Fock space consists of vectors of the type 
\begin{equation}
\left| N\left( k\right) -m_{i},m_{i}\right\rangle =\prod_{i}\sqrt{\frac{%
\left( N\left( k_{i}\right) -m_{i}\right) !}{N\left( k_{i}\right) !m_{i}!}}
\left( b^{+}\left( k_{i}\right) \right) ^{m_{i}}\left| N\left( k\right)
,0\right\rangle ,
\end{equation}
where $m_{i}=0,1,...,N\left( k\right) $.

We interpret the function $N\left( k\right) $ as a geometric characteristic
of the momentum space which has formed during the quantum period in the
evolution of the Universe, when topology transformations took place. Then
assigning a specific value for the function $N\left( k\right) $, expressions
(\ref{d}) and (\ref{f}) define the ground state for respective particles.

\section{Origin of the spectral number of fields}

A rigorous derivation of properties of the function $N\left( k\right) $
requires studying processes involving topology changes which took place
during the quantum stage of the evolution of the Universe. At the moment, we
do not have an exact model describing the formation of $N\left( k\right) $
and, therefore, our consideration will have a phenomenological character. We
assume that upon the quantum period of the Universe the matter was
thermalized with a very high temperature. Then, as the temperature dropped
during the early stage of the evolution of the Universe, the topological
structure of the space (and the spectral number of fields) has tempered and
the subsequent evolution resulted only in the cosmological shift of the
physical scales.

There exist at least two possibilities. The first and the simplest
possibility is the case when a transformation in the topology of space
results in an equal transformation of the number of modes for all fields
(regardless of the type of the fields). In this case processes involving
topology changes generate a unique function $N\left( k\right) $ which is the
same for all fields. This case was considered in Ref. \cite{KT02}. However,
the mathematical structure of MOFT reserves a more general possibility when
the formation of the spectral distribution of modes goes in independent ways
for different fields. In this case every particular field $\psi _{a}$ will
be characterized by its own function $N_{a}\left( k\right) $ which can
possess specific features. Which case is realized in the nature can be
determined only by means of confrontation with observations and below we
consider both cases.

Upon the quantum period the Universe is supposed to be described by the
homogeneous metric of the form 
\begin{equation}
ds^{2}=dt^{2}-a^{2}\left( t\right) dl^{2},
\end{equation}
where $a\left( t\right) $ is the scale factor, and $dl^{2}$ is the spatial
interval. It is expected that the state of fields was thermalized with a
very high temperature $T>T_{Pl}$ where $T_{Pl}$ is the Planck temperature.
Then the state of fields was characterized by the thermal density matrix
with mean values for occupation numbers 
\begin{equation}
\left\langle N\left( k,n\right) \right\rangle =\left( \exp \left( \frac{
n\omega _{k}-\mu _{k}}{T}\right) \pm 1\right) ^{-1},  \label{on}
\end{equation}
where the signs $+$ and $-$ corresponds to bosonic and fermionic fields
respectively and the chemical potential $\mu _{k}$ can be expressed via the
spectral number of fields as 
\begin{equation}
N\left( k\right) =\sum_{n}\left( \exp \left( \frac{n\omega _{k}-\mu _{k}}{T}
\right) \pm 1\right) ^{-1}.  \label{N}
\end{equation}
In particular, from (\ref{N}) we find that for fermions in the region $%
\omega _{k}-\mu _{k}\ll T$ the relation between the spectral number of
fields and $\mu _{k}$ takes the form 
\begin{equation}
N\left( k\right) \sim T/\left( -\mu _{k}\right) +T/\left( \omega _{k}-\mu
_{k}\right) .
\end{equation}

Consider now the first case when the spectral number of fields $N\left(
k\right) $ is a unique function for all fields. It is well known that near
the singularity the evolution of the Universe is governed by a scalar field
(responsible for a subsequent inflationary phase), while all other fields
can be neglected. We assume that the same field is responsible for topology
transformation processes which took place in the early Universe. Thus, we
can expect that the state of the scalar field was characterized by the
thermal density matrix (\ref{on}) with $\mu =0$ (for the number of fields
varies). On the early stage $m\ll T$, and the temperature and the energy of
scalar particles depend on time as $T=\widetilde{T}/a\left( t\right) $, $k= 
\widetilde{k}/a\left( t\right) $. When the temperature drops below a
critical value $T_{\ast }$ , which corresponds to the moment $t_{\ast }\sim
t_{pl}$, topological structure (and the number of fields) tempers. This
generates the value of the chemical potential for scalar particles $\mu \sim
T_{\ast }$.

Let us neglect the temperature corrections, which are essential only at $%
t\sim t_{\ast }$ and whose role is in smoothing the real distribution $N_{k}$
. Then at the moment $t\sim t_{\ast }$ the ground state of the scalar field
will be described by (\ref{d}) with $\mu _{k}=\mu =const\sim T_{\ast }$.
During the subsequent evolution the\ physical scales are subjected to the
cosmological shift, however the form of this distribution in the commoving
frame must remain the same. Thus, on the later stages $t\geq t_{\ast }$, we
find 
\begin{equation}
N\left( k\right) =1+\left[ \frac{\widetilde{k}_{1}}{\Omega _{k}\left(
t\right) }\right] ,\,\,  \label{NN}
\end{equation}
where $\Omega _{k}\left( t\right) =\sqrt{a^{2}\left( t\right) k^{2}+%
\widetilde{k}_{2}^{2}}$, $\widetilde{k}_{1}\sim a_{0}\mu $, and $\widetilde{k%
}_{2}\sim a_{0}m$ ($a_{0}=a\left( t_{\ast }\right) $). From (\ref{NN}) we
see that there is a finite interval of wave numbers $k\in \lbrack k_{\min
}\left( t\right) ,k_{\max }\left( t\right) ]$ on which the number of fields $%
N_{k}$ changes its value from $N_{k}=1$ (at the point $k_{\max }$) to the
maximal value $N_{\max }=1+\left[ \widetilde{k}_{1}/\widetilde{k}_{2}\right] 
$ (at the point $k_{\min }$) where the boundary points of the interval of $k$
depend on time and are expressed via the free phenomenological parameters $%
\widetilde{k}_{1}$ and $\widetilde{k}_{2}$ as follows 
\[
k_{\max }=\frac{1}{a\left( t\right) }\sqrt{\widetilde{k}_{1}^{2}-\widetilde{k%
}_{2}^{2}},\;\;k_{\min }=\frac{1}{a\left( t\right) }\sqrt{\widetilde{k}%
_{1}^{2}/\left( N_{\max }-1\right) ^{2}-\widetilde{k}_{2}^{2}}.
\]
Out of this interval the number of fields remains constant i.e., $N\left(
k\right) =N_{\max }$ for the range $k\leq k_{\min }\left( t\right) $ and $%
N\left( k\right) =1$ for the range $k\geq k_{\max }\left( t\right) $. From
restrictions on parameters of inflationary scenarios we get $m\lesssim
10^{-5}m_{Pl}$ \ which gives $N_{\max }\gtrsim 10^{5}T_{\ast }/m_{pl}$,
where $T_{\ast }$ is the critical temperature at which topology has been
tempered. Thus, substituting (\ref{NN}) in (\ref{N}) we define the values of
the chemical potentials $\mu _{k}$ for all other particles.

Consider now the second case when the spectral number of modes $N\left(
k\right) $ forms independently for different fields. In this case bosonic
fields are described by the same distribution (\ref{NN}) in which , however,
the parameters $\widetilde{k}_{1}$ and $\widetilde{k}_{2}$ are free
phenomenological parameters which are specific for every particular field.
In particular, for massless fields we find $\widetilde{k}_{2}=k_{\min }=0$
and $N\left( k\right) =1+\left[ k_{\max }/k\right] $. In the case of
fermions the chemical potential $\mu _{k}$ cannot vanish and for $T\geq
T_{\ast }$ it should take some rest value $\mu _{k}=\epsilon _{0}$
(otherwise $N\left( k,0\right) \rightarrow \infty $). Thus, in the same way
as in the case of bosons, we find $N\left( k\right) =1+\theta \left( k_{\max
}-k\right) $, where $\theta \left( x\right) $ is the Heaviside step function
and $k_{\max }\sim T_{\ast }a\left( t_{\ast }\right) /a\left( t\right) $. In
this case the spectral number of fermions is characterized by the only
phenomenological parameter and $N\left( k\right) =N_{\max }=2$ as $k<k_{\max
}$.

In this manner we have shown that properties of the spectral number of
fields can be different, depending on which case is realized in the nature.
However, we note that if in the first case the spectral number of fields $%
N\left( k\right) $ can be considered as a new geometric characteristics
which straightforwardly defines properties of space and hence of all matter
fields, in the second case we, rigorously speaking, cannot use such an
interpretation. Moreover, if the last case is really realized in the nature,
it should relate to yet unknown processes. Therefore, in the next section we
will discuss the first possibility only.

In conclusion of this section we point out to a formal analogy between MOFT
and the Hagedorn theory \cite{Hag}. This analogy, however, is not complete,
for in MOFT the number of sorts of particles is always finite and the total
energy density $\varepsilon $ behaves like in the standard theory, e.g., for 
$T\gg m$ we can show that $\varepsilon =\varkappa \left( T\right) T^{4}$,
where $\varkappa \rightarrow const$ as $T\rightarrow \infty $. We also note
that the real distribution can be different from (\ref{NN}), which depends
on the specific picture of topology transformations in the early Universe
and requires the construction of the exact theory (in particular, thermal
corrections smoothen the step-like distribution (\ref{NN})). However we
believe that the general features of $N_{k}$ will remain the same.

\section{Effective dimension of the Universe}

The growth of the spectral number of fields which takes place in the range
of wave numbers $k_{\min }\left( t\right) <k<k_{\max }\left( t\right) $
leads to the fact that in this range our Universe has to demonstrate
nontrivial geometric properties. Indeed, it was recently shown that in the
same range of scales a stable equilibrium distribution of baryons requires a
fractal behavior with dimension $D\approx 2$ \cite{K02}, while the Newton's
and Coulomb's energies of interaction between point particles show the
logarithmic behavior \cite{KT02}. We recall that the logarithmic potential $%
\ln \left( r\right) $ gives the solution of the Poisson equation with a
point source for two dimensions. Both these phenomena are in agreement with
the observed picture of the Universe and it turns out that they have a
common pure geometrical interpretation. Namely, we can say that in the range
of scales $r_{\min }<r<r_{\max }$ (where $r_{\min }=2\pi /k_{\max }$) some
kind of reduction of the dimension of space happens.

Indeed, the simplest way to demonstrate this is to compare the spectral
number of modes in the interval between $k$ and $k+dk$ in MOFT, which is
given by the measure (\ref{mes}) (i.e., $N\left( k\right) d^{3}k/\left( 2\pi
\right) ^{3}$), with the spectral number of modes for $n$ dimensions in the
standard field theory (i.e., $d^{n}k/\left( 2\pi \right) ^{n}$). Hence we
can define the effective dimension $D$ of space as follows 
\begin{equation}
k^{3}N\left( k\right) \sim k^{D}.  \label{dim}
\end{equation}
In the standard field theory $N\left( k\right) =1$ and we get $D=3$, while
in MOFT the properties of the function $N\left( k\right) $ were formed
during the quantum period in the evolution of the Universe and depend on
specific features of topology transformation processes. Thus, in general
case, the effective number of dimensions $D$ may\ take different values for
different intervals of scales. If we take the value (\ref{NN}) we find that
in the range of wave numbers $k_{\max }\geq k\geq k_{\min }$ (where the
function $N\left( k\right) $ can be approximated by $N\left( k\right) \sim
k_{\max }/k$) the effective dimension of space is indeed $D\approx 2$. The
scale $r_{\min }=2\pi /k_{\max }$ can be called an effective scale of
compactification of $3-D$ dimensions, while the scale $r_{\max }$, if it
really exists, represents the boundary after which the dimension $D=3$
restores. We note also that after this scale the fractal picture of the
Universe crosses over to homogeneity and the standard Newton's law restores.

In MOFT, $N\left( k\right) $ represents the operator of the number of fields
( \ref{N(k)}) which is common for all types of fields and, therefore, plays
the role of the density operator for the momentum space. We note that $%
N\left( k\right) $ is an ordinary function only in the case when topology
changes are suppressed. In the coordinate representation the number of
fields is described by an operator $N\left( x\right) $ (e.g., see Ref. \cite
{KT02}) which defines the density of physical space, i.e., the volume
element is given by $dV=N\left( x\right) d^{3}x$.

Consider the relation between these two operators. In what follows we, for
the sake of convenience, consider a box of the length $L$ and will use the
periodic boundary conditions (i.e., ${\bf k}=2\pi {\bf n}/L$ and, as $%
L\rightarrow \infty $, $\sum_{k}\rightarrow \int \left( L/2\pi \right)
^{3}d^{3}k$). From the dynamical point of view the operator $N\left(
k\right) $ has a canonically conjugated variable $\vartheta \left( k\right) $
such that 
\begin{equation}
\left[ \vartheta \left( k\right) ,N\left( k^{\prime }\right) \right]
=\vartheta \left( k\right) N\left( k^{\prime }\right) -N\left( k^{\prime
}\right) \vartheta \left( k\right) =i\delta _{k,k^{\prime }}\,.
\end{equation}
These two operators can be used to define a new set of creation and
annihilation operators 
\begin{equation}
\Psi ^{+}\left( k\right) =\sqrt{N\left( k\right) }e^{i\vartheta \left(
k\right) },\Psi \left( k\right) =e^{-i\vartheta \left( k\right) }\sqrt{
N\left( k\right) }
\end{equation}
which obey the standard commutation relations 
\begin{equation}
\left[ \Psi \left( k\right) ,\Psi ^{+}\left( k^{\prime }\right) \right]
=\delta _{k,k^{\prime }}
\end{equation}
and have the meaning of the creation/annihilation operators for field modes,
e.g., the density operator for the momentum space is defined simply as $%
N\left( k\right) =\Psi ^{+}\left( k\right) \Psi \left( k\right) $. In the
case when topology transformations are suppressed the operator $\Psi $ can
be considered as a classical scalar field $\varphi $ which characterizes the
density of physical space\footnote{%
We note that if every field is characterized by its own spectral number
density $N\left( k\right) $, we have to introduce a set of independent
classical scalar fields (one for every quantum field). This gives a good
chance to explain origin of Higgs fields in particle theory.} and, in
general, depends on time and space coordinates (in what follows we consider
the spatial dependence only). Indeed, in applying to the ground state $\Phi
_{0}=\left| N\left( k\right) \right\rangle $ (which is expressed by the
occupation numbers (\ref{d}) and ( \ref{f})) the operators $\Psi $ and $\Psi
^{+}$ change the number of modes by one, while the total number of modes $%
N\rightarrow \infty $. In this sense the state $\Phi _{0}^{\prime }=\Psi
\Phi _{0}\approx \Phi _{0}$. Thus, the classical field $\varphi $ may be
defined by relations 
\begin{equation}
\varphi \left( k\right) =\left\langle N\left( k\right) \left| \Psi \left(
k\right) \right| N\left( k\right) +1\right\rangle ,\;\varphi ^{\ast }\left(
k\right) =\left\langle N\left( k\right) +1\left| \Psi ^{+}\left( k\right)
\right| N\left( k\right) \right\rangle ,
\end{equation}
which gives $\varphi \left( k\right) =\varphi ^{\ast }\left( k\right) =\sqrt{
N\left( k\right) }$. The coordinate dependence of this field can be found in
the standard way (e.g., see Ref. \cite{Landau}) 
\begin{equation}
\Psi \left( {\bf r}\right) =e^{-i{\bf r}\widehat{{\bf P}}}\Psi \left(
0\right) e^{i{\bf r}\widehat{{\bf P}}}
\end{equation}
where $\widehat{{\bf P}}$ is the total momentum operator. If in states $%
N_{i} $ and $N_{f}$ the system possesses fixed momenta ${\bf P}_{i}$ and $%
{\bf P} _{f}$, then 
\begin{equation}
\left\langle N_{i}\left| \Psi \left( {\bf r}\right) \right|
N_{f}\right\rangle =\exp \left( -i{\bf rk}_{if}\right) \left\langle
N_{i}\left| \Psi \left( 0\right) \right| N_{f}\right\rangle  \label{x}
\end{equation}
where ${\bf k}_{if}={\bf P}_{i}-{\bf P}_{f}$. We note that the
creation/annihilation of a single mode is accompanied with the
increase/decrease in the number of ''dark'' particles (\ref{hp}) by $N\left(
k\right) $ and hence in the total momentum by ${\bf P}_{k}=zN\left( k\right) 
{\bf k}$, where $z$ is the number of different types of bosonic fields.
Thus, from (\ref{x}) we find that coordinate dependence will be described by
the sum of the type 
\begin{equation}
\varphi \left( x\right) \sim \sum_{k}c_{k}\exp \left( -i{\bf P}_{k}{\bf r}
\right) .  \label{s}
\end{equation}
where $c_{k}\sim \sqrt{N\left( k\right) }$. In the range of wave numbers $%
k_{\max }\geq k\gg k_{\min }$ we can use the approximation $N\left( k\right)
\simeq 1+k_{\max }/k$ and let the length of the box be $L<r_{\max }=2\pi
/k_{\min }$. Then we find that in the sum (\ref{s}) the maximal wavelength
is indeed restricted by 
\begin{equation}
\ell \leq \ell _{0}<L,
\end{equation}
where $\ell _{0}=2\pi /zk_{\max }=r_{\min }/z$, which means that at least in
one direction the space box is effectively compactified to the size $\ell
_{0}$. From the physical standpoint such a compactification will be
displayed in irregularities of the function $N\left( x\right) \sim \left|
\varphi \left( x\right) \right| ^{2}$ (we point out that the time dependence
will randomize phases in (\ref{s})). E.g., the function $N\left( x\right) $
may take essential values $N\left( x\right) \sim <N>$ only on thin two
dimensional surfaces of the width $\sim \ell _{0}$ and rapidly decay
outside, which may explain the formation of the observed fractal
distribution of galaxies and the logarithmic behavior of the Newton's
potential. In considering the box of the larger and larger size $L$ $\gg
r_{\max }$we find that $N\left( k\right) \rightarrow N_{\max }\sim r_{\max
}/r_{\min }$ and the effect of the compactification disappears (no
restrictions on possible values of wavelengths emerge). Thus at scales $\ell
>r_{\max }$ dimension $D=3$ restores which restores the standard Newton's
law and the distribution of galaxies crosses over to homogeneity.

\section{Conclusions}

In this manner we have shown that MOFT predicts a rather interesting physics
for the range of scales $r_{\min }<r<$ $r_{\max }$. First of all we point
out to the fact that in this range the Universe acquires features of a
two-dimensional space whose distribution in the observed 3-dimensional
volume has an irregular character. This can provide a natural explanation to
the observed fractal distribution of galaxies and the dark matter problem.
We note that such properties originate from a primordial thermodynamically
equilibrium state and are in agreement with the homogeneity and isotropy of
the Universe. Thus, such a picture of the Universe represents a homogeneous
background, while gravitational potential fluctuations should be considered
in the same way as in the standard model. Nevertheless, we can state that in
the range of wavelengths $r_{\min }<\lambda <$ $r_{\max }$ the propagation
of perturbations will correspond to the 2-dimensional law.

The two-dimensional character of the Universe in the range of scales means
that we should expect that the standard Newtonian kinematics breaks also
down. Some realization of such a possibility is contained in MOND (the
Modified Newtonian Dynamics) proposed by Milgrom in Ref. \cite{mil}. We
believe that MOFT provides a rigorous way to derive such a modification
which, however, requires a separate investigation.

In the present paper we have supposed that upon the quantum period in the
evolution of the Universe the spectral number of fields $N\left( k\right) $
is conserved, which means that $N\left( k\right) $ depends on time via only
the cosmological shift of scales (i.e., $k\left( t\right) \sim 1/a\left(
t\right) $, where $a\left( t\right) $ is the scale factor). However, there
are reasonable arguments which show that the number of fields should
eventually decay. Indeed, if we take into account thermal corrections to $%
N\left( k\right) $, then instead of (\ref{NN}) we get the expression of the
type $N\left( k\right) \sim 1+T_{\ast }/k+...$. In this case the temperature 
$T_{\ast }$ plays the role of the maximal wave number $k_{\max }$. Why at
present this temperature is so small $T_{\ast }\ll T_{\gamma }$ (where $%
T_{\gamma }$ is the CMB temperature) requires a separate explanation. The
situation is different when we assume that the number of modes may decay. At
the moment, we do not have an exact model describing the decay of fields,
however, phenomenologically the decay can be described by the expression of
the type $N\left( k\right) \sim N_{0}\left( k\right) e^{-\Gamma _{k}t}$
(where $\Gamma _{k}$ is the period of the half-decay which, in general, can
depend on wave numbers $k$). In the simplest case $\Gamma _{k}=const$ and
this would lead to an additional monotonic increase of the minimal scale $%
r_{\min }\sim a\left( t\right) e^{\Gamma t}$, while the maximal scale
changes according to the cosmological shift only $r_{\max }\sim a\left(
t\right) $. We note that the ground state of fields $\Phi _{0}$ contains
hidden bosons (\ref{hp}). The decay of modes transforms them into real
particles and, therefore, the decay is accompanied by an additional
reheating which should change the temperature of the primordial plasma
(which can explain the difference between $T_{\ast }$ and $T_{\gamma }$).
Besides, the additional increase in $r_{\min }$ means that the number of
baryons contained within a radius $r_{\min }$ depends on time $N_{b}\left(
r_{\min }\right) \sim e^{3\Gamma t}$ . Thus, on early stages of the
evolution of the Universe we find that $N_{b}\left( r_{\min }\right) $ $\ll 1
$ and the stable equilibrium distribution of baryons had the fractal
character $N_{b}\left( r\right) \sim r^{2}$ for all scales $r<r_{\max }$.
Eventually, $r_{\min }$ increases, the fractal distribution of baryons below
the scale $r_{\min }$ becomes unstable, and the larger and larger number of
baryons $N_{b}\left( r_{\min }\right) \sim e^{2\Gamma t}$ are involved into
the structure formation. In such a picture the smallest objects appear first
and then they form groups, galaxies, clusters, etc.. The advantage of such a
scenario is the fact that it does not require the existence of primeval
perturbations of the metric.

We also point out to the violation of the Pauli principle for wavelengths $%
\lambda >r_{\min }$ (more than one fermion can have a wavelength $\lambda $
). Such particles are located in the volume $\gtrsim r_{\min }^{3}$ and at
laboratory scales the portion of states violating the statistics is
extremely suppressed $P\lesssim \left( L/r_{\min }\right) ^{3}$, where $L$
is a characteristic spatial scale of a system under measurement (e.g., if as
such a scale $L$ we take the Earth radius, this factor will be still
extremely small $P\sim 10^{-32}$). We note that for a system of particles
which is in the thermal equilibrium state the characteristic scale $L\sim
1/T $ (where $T$ is the temperature). Thus, if the number of fields is
conserved, MOFT will not change the predictions of the standard model of
nucleosynthesis, e.g., corrections will have the factor $P\sim \left(
1/T_{\gamma }r_{\min }\right) ^{3}$ which does not depend on time. However,
the nucleosynthesis can introduce some restrictions on the possible rate of
the decay of modes $\Gamma $. This problem, however, requires the further
investigation.

In conclusion we point out to the fact that, according to MOFT, the
violation of the Pauli principle serves as an indicator of a nontrivial
topology of space and, therefore, the possibility to detect such a violation
in laboratory experiments represents an extremely intriguing problem.

\end{document}